\documentclass[aps,prb,twocolumn,showpacs,amssymb,amsmath]{revtex4}
\usepackage{graphics,epsfig}

\begin{document}
\preprint{SNUTP 02-005}

\title{Dynamic transition and resonance in current-driven arrays of
Josephson junctions}
\author{Gun Sang Jeon}
\affiliation
{Center for Strongly Correlated Materials Research, Seoul National University,
Seoul 151-747, Korea}
\author{Jong Soo Lim}
\author{Hyun Jin Kim}
\author{M.Y. Choi}
\affiliation{Department of Physics, Seoul National University,
Seoul 151-747, Korea}

\begin{abstract}
We consider a two-dimensional fully frustrated Josephson-junction array,
which is driven uniformly by oscillating currents.
As the temperature is lowered, there emerges a dynamic phase transition to an ordered
state with nonzero dynamic order parameter for small currents.
The transition temperature decreases monotonically with the driving amplitude,
approaching zero at a certain critical value of the amplitude.
Above the critical value, the disordered phase and the dynamically ordered phase are
observed to appear alternatively.
The characteristic stochastic resonance behavior of the system is also examined,
which reveals that the resonance behavior of odd and even harmonics can be different
according to the zero-temperature state.
\end{abstract}

\maketitle

\section{Introduction}

Dynamic responses of cooperatively interacting many-body systems to time driven
perturbations are extremely important technologically and involve intriguing physics,
resulting in intensive investigation in recent years.
Among the well-known systems is the simple kinetic Ising model under oscillating
magnetic fields.\cite{tome,Korniss,SR}
It has been revealed that spontaneous symmetry breaking takes place at a finite
strength of the oscillating field\cite{tome,Korniss} and that in two dimensions the transition belongs
to the same universality class as the equilibrium two-dimensional (2D)
Ising model.\cite{Korniss}
The stochastic resonance (SR) phenomena have also been studied.\cite{SR}

Another interesting example is the dynamics of the fully frustrated Josephson-junction
array (FFJJA), which possesses the same Z$_2$ symmetry as the Ising model in addition
to the continuous U(1) symmetry.  Unlike the Ising model, the FFJJA has real intrinsic
dynamics derived from the Josephson relations and thus grants direct experimental
realizations.  There have been some studies on the dynamic properties of the
FFJJA \cite{mon,luo} and recently dynamic transitions have also been investigated
in the FFJJA driven by uniform dc currents \cite{marconi} or staggered oscillating
magnetic fields.\cite{gun}
Similarly to the kinetic Ising model, the FFJJA driven by staggered oscillating
fields has been shown to exhibit a dynamic phase transition and an SR behavior.
In the presence of weak fields, the dynamic phase transition belongs to the same
universality class as the equilibrium Z$_2$ transition in the fully frustrated $XY$
(FF$XY$) model.  At strong fields, in contrast, a different universality class has
been suggested.

In this paper, we investigate the FFJJA in the presence of uniform oscillating currents,
which is easier to realize in experiment than the system under staggered oscillating
magnetic fields.  The dynamic properties of the system is examined with attention
paid to the dynamic transition and the SR phenomena.  At low temperatures the
chirality in the FFJJA displays antiferromagnetic ordering.  To describe the dynamic
transition associated with the antiferromagnetic ordering of the chirality,
we introduce the dynamic order parameter, which is the staggered magnetization
averaged over one period of the uniform oscillating current, and investigate its
behavior with respect to the amplitude and the frequency of the driving current as
well as to the temperature.  At zero temperature, as the driving amplitude is raised,
the dynamically ordered phase and the disordered one are observed to appear
alternatively.  As the temperature is increased, the ordered phase undergoes
a dynamic phase transition to the disordered phase.
Through the use of the finite-size scaling analysis, the dynamic transition
is shown to belong to the same universality class as the equilibrium Z$_2$ transition
present in the FF$XY$.
We also examine the SR phenomena in the system.
As manifested by the signal-to-noise ratio (SNR) at several harmonics,
odd and even harmonics are revealed to exhibit different SR behaviors.
It is discussed in view of the variation of the zero-temperature states with
the driving current amplitude.

This paper consists of five sections: Section II introduces the fully frustrated
Josephson-junction array, driven uniformly by alternating currents,
together with the dynamic order parameter, describing the dynamic transition of
the system, and presents the zero-temperature behavior.
In Sec. III the behavior of the dynamic order parameter at finite temperatures
is investigated in detail, on the basis of which the dynamic phase diagram is
constructed on the plane of the temperature and the driving amplitude.
The nature of the transition is also examined and the equilibrium Z$_2$ universality
class is identified.
Section IV is devoted to the power spectrum of the staggered magnetization
and the corresponding signal-to-noise ratio, which reveals the characteristic
resonance behavior of the system with the appropriate values of the parameters.
Finally, a brief summary is given in Sec. V.

\section{Fully Frustrated Array Driven by Alternating Current}

To begin with, we consider the equations of motion for phase angles \{$\phi_i$\} of
the superconducting order parameters in grains forming an $L \times L$ square lattice
with unit lattice constant.  Within the resistively-shunted-junction model
under the fluctuating twist boundary conditions,\cite{bjk} they read:
\begin{equation}
{\sum_j}' \left[ \frac{d{\widetilde{\phi}}_{ij}}{dt} +
\sin({\widetilde{\phi}}_{ij} - {\bf r}_{ij}\cdot{\bf \Delta}) + \eta_{ij} \right] = 0,
\label{e1}
\end{equation}
where the primed summation runs over the nearest neighbors of
grain $i$ and the thermal noise current $\eta_{ij}$ satisfies
$\langle \eta_{ij}(t) \eta_{kl}(t') \rangle =
2T\delta(t-t')(\delta_{ik}\delta_{jl} - \delta_{il}\delta_{jk})$
at temperature $T$. We have used the abbreviation
${\widetilde{\phi}}_{ij} \equiv \phi_i - \phi_j - A_{ij}$ and
${\bf r}_{ij} \equiv {\bf r}_i - {\bf r}_j$ with ${\bf r}_i =
(x_i, y_i)$ denoting the position of grain $i$. Note that ${\bf
r}_{ij}$ for nearest neighboring grains is a unit vector since the
lattice constant has been set equal to unity. We have also written
the energy and the time in units of $\hbar i_c /2e$ and $\hbar
/2eRi_c$, respectively, with the critical current $i_c$ and the
shunt resistance $R$. The dynamics of the twist variables ${\bf
\Delta} \equiv (\Delta_x, \Delta_y)$, which are included in the
fluctuating twist boundary conditions to allow fluctuations in the
phase difference across the whole system, is governed by the
equations
\begin{align}
& \frac{d\Delta_x}{dt}
  = \frac{1}{L^2} \sum_{{\langle ij \rangle}_x}\sin({\widetilde{\phi}}_{ij} - \Delta_x)
    + \eta_{\Delta_x} - I_0\sin(\Omega t), \nonumber \\
& \frac{d\Delta_y}{dt}
  = \frac{1}{L^2} \sum_{{\langle ij \rangle}_y}\sin({\widetilde{\phi}}_{ij} - \Delta_y)
    + \eta_{\Delta_y},
\label{e2}
\end{align}
where $\sum_{\left\langle ij \right\rangle_a}$ denotes the summation over
all nearest neighboring pairs in the $a \,(= x, y)$ direction, $\eta_{\Delta_a}$ satisfies
$\left\langle \eta_{\Delta_a}(t+\tau)\eta_{\Delta_{a'}}(t) \right\rangle
= (2T/L^2)\delta(\tau)\delta_{a,a'}$,
and the oscillating current $I_0 \sin(\Omega t)$ is injected in the $x$ direction.
In the Landau gauge, the bond angle $A_{ij}$, given by the line integral of
the vector potential, takes the values
\begin{equation*}
A_{ij} =
\begin{cases}
0       & \textrm{for ${\bf r}_j = {\bf r}_i + {\bf \hat{x}}$} \\
\pi x_i & \textrm{for ${\bf r}_j = {\bf r}_i + {\bf \hat{y}}$}
\end{cases}.
\end{equation*}

For the study of the transition associated with the ${\rm Z}_2$ symmetry in the FFJJA,
it is convenient to consider the chirality
\begin{equation}
C({\bf R},t) \equiv {\rm sgn}
\left[ \sum_{\bf P} \sin \left( \widetilde{\phi}_{ij}(t)
      - {\bf r}_{ij}\cdot {\bf \Delta}(t) \right) \right]
\end{equation}
and the staggered magnetization
\begin{equation}
m(t) \equiv \frac{1}{L^2} \sum_{\bf R} (-1)^{x_i + y_i} C({\bf R},t),
\end{equation}
where $ \sum_{\bf P}$ denotes the directional plaquette summation of links
around the dual lattice site
${\bf R} \equiv {\bf r}_i + (1/2)(\hat{\bf x}+\hat{\bf y})$.
To probe the dynamic transition in the presence of an oscillating uniform current,
we define the dynamic order parameter as the staggered magnetization averaged over
one period of the oscillating current
\begin{equation}
Q \equiv \frac{\Omega}{2\pi} \left| \oint m(t) dt \right|.
\end{equation}

In the numerical calculation, the sets of the equations of motion in Eqs.~(\ref{e1})
and (\ref{e2}) are integrated via the modified Euler method with time steps of size
$\Delta t = 0.05$.  We have varied the step size, only to find no essential difference.
Typically, data have been averaged over 5000 driving periods after the data obtained
from the initial 1000 periods discarded;
the appropriate stationarity has been verified.
In addition, five independent runs have been performed, over which the average
has also been taken, and systems of size up to $L = 32$ have been considered.

\begin{figure}
\epsfig{width=0.4\textwidth,file=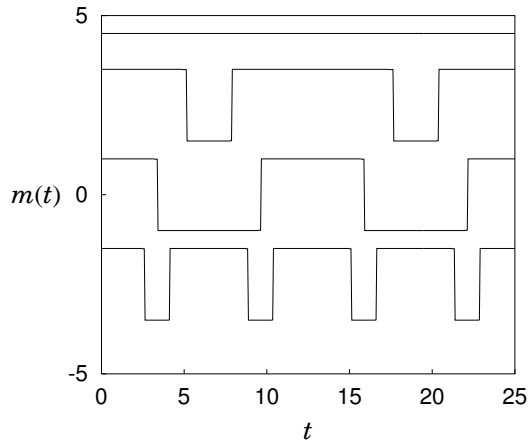}
\vspace{0.3cm}
\caption[Time evolution of the staggered magnetization at zero temperature]
{Time evolution of the staggered magnetization at zero temperature for driving
frequency $\Omega/2\pi = 0.08$ and amplitudes $I_0 = 0.98, 1.03, 1.50$, and $2.0$
from above.  For clarity, the data for $I_0 = 0.98$ and $1.03$ have been shifted
upward by $3.5$ and $2.5$, respectively, whereas those for $I_0 = 2.0$ shifted
downward by $2.5$.}
\label{fig1}
\end{figure}

\begin{figure}
\epsfig{width=0.4\textwidth,file=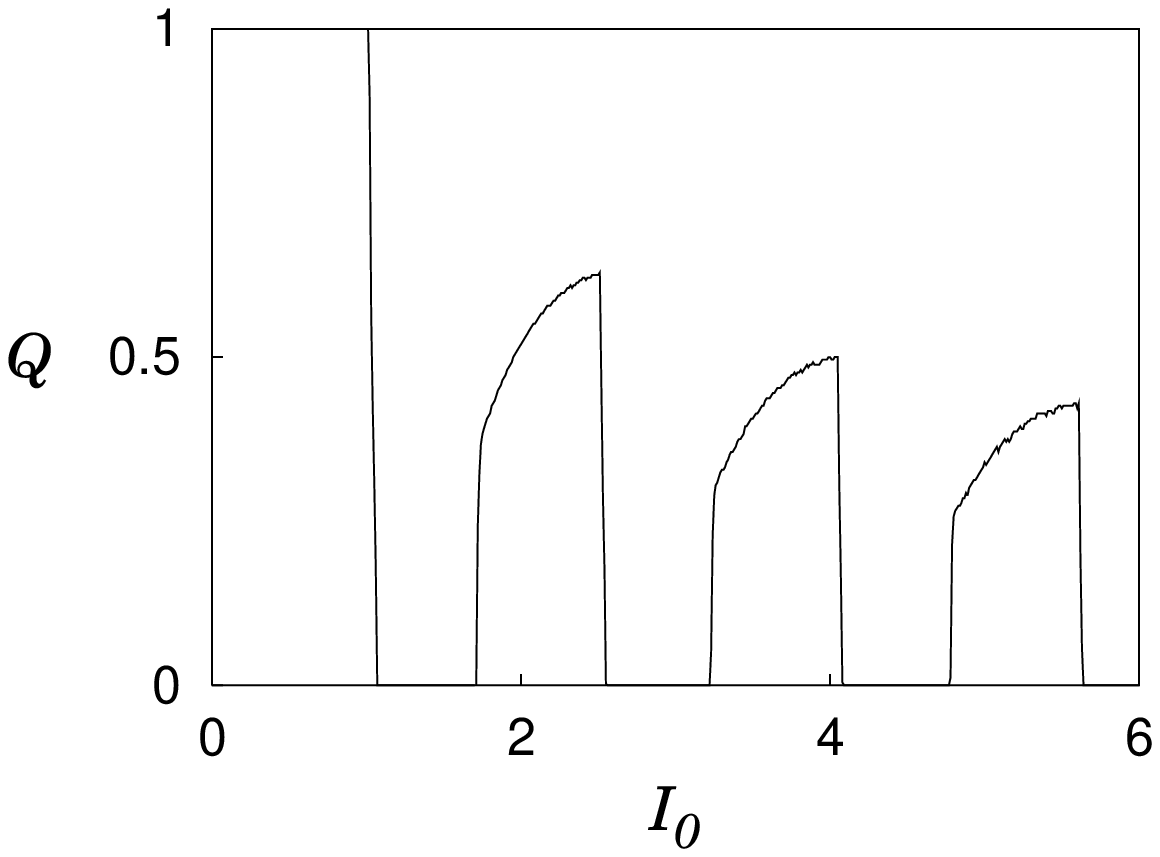} \\
\vspace{0.1cm}
(a) \\
\epsfig{width=0.4\textwidth,file=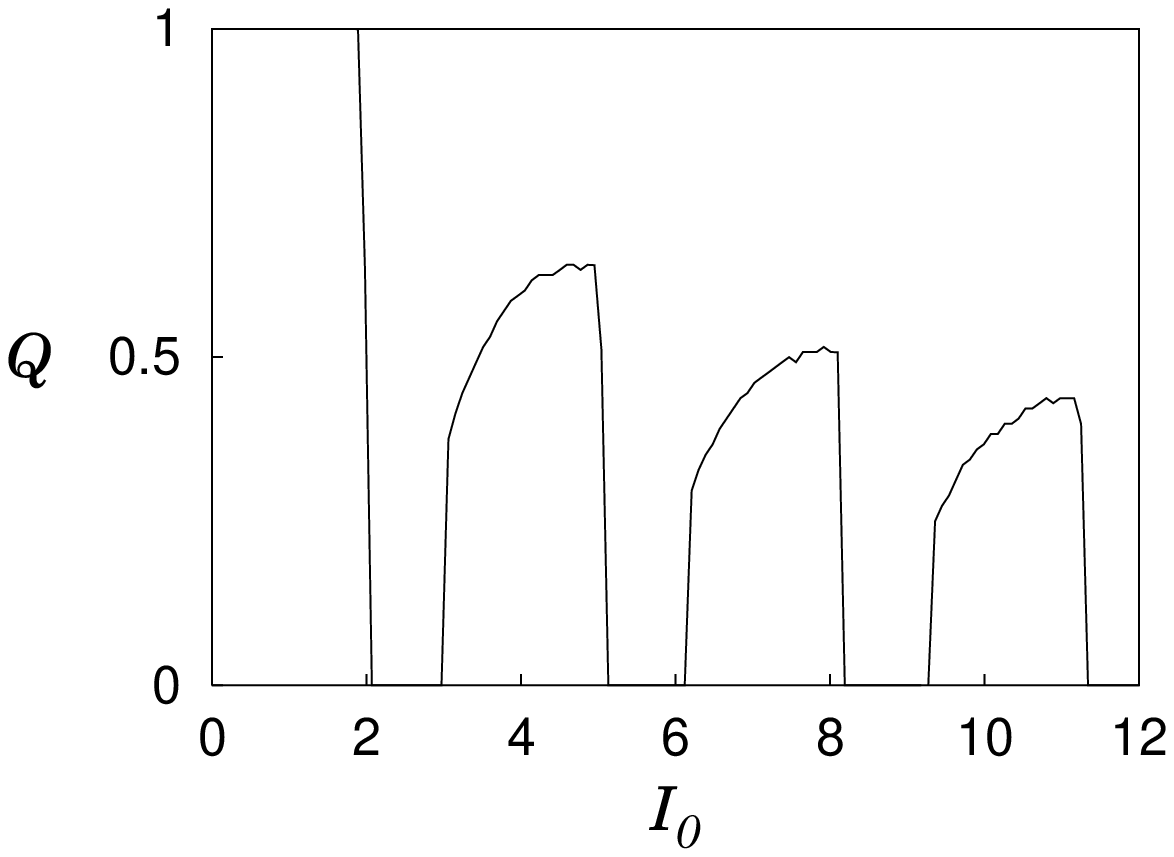} \\
\vspace{0.1cm}
(b)
\vspace{0.3cm}
\caption[Dynamic order parameter as a function of the driving amplitude at zero
temperature]
{Dynamic order parameter as a function of the driving amplitude at zero temperature
in the system of size $L =16$ with (a) $\Omega/2\pi = 0.08$; (b) $\Omega/2\pi = 0.16$.}
\label{fig2}
\end{figure}

In the FFJJA, it is well known that there are two kinds of
antiferromagnetic chirality ordering at zero temperature. The
injection of uniform oscillating currents plays the role of
tilting sinusoidally the 2D ``egg-carton" lattice pinning
potential. Accordingly, if the driving amplitude $I_0$ of the
injected currents is large enough to overcome the lattice pinning
potential, oscillations between the two ground states are expected
to take place. In Fig.~\ref{fig1}, we display the time evolution
of the staggered magnetization $m(t)$ at zero temperature, evolved
from the initial condition $m(t{=}0) = 1$ for driving frequency
$\Omega/2\pi = 0.08$. For small $I_0$, the system is shown to stay
in the state with $m(t) =1$ in spite of the driving current. When
the amplitude exceeds the value ${I_0}^{\left( 0 \right)} \approx
1.01$, the system is driven out of the $m = 1$ state and the
staggered magnetization oscillates between $m = \pm 1$; this
arises since the chirality lattice moves collectively over the
lattice potential under large current driving.  In case that the
driving induces the chirality lattice to move over only a single
lattice barrier in the first half of the period, the system
oscillates between the two ground states symmetrically (see the
staggered magnetization for $I_0=1.50$ in Fig.~\ref{fig1}). On the
other hand, further increase of $I_0$ allows the chirality lattice
to go over another barrier and yields asymmetric oscillations of
the system between the ground states, as demonstrated by the
staggered magnetization for $I_0=2.0$ in Fig.~\ref{fig1}.  The
symmetry is restored and broken alternatively with the increase of
$I_0$, which is manifested by the zero-temperature behavior of $Q$
with $I_0$ in Fig.~\ref{fig2}.  As expected in Fig.~\ref{fig1},
the dynamic order parameter $Q$, which is unity below
${I_0}^{\left( 0 \right)}$, is shown to decrease abruptly to zero
for $I > {I_0}^{\left( 0 \right)}$. Further increase of $I_0$
reveals that the dynamically ordered state ($Q > 0$) and the
disordered one ($Q =0$) appear alternatively in the system. It is
observed that the saturation value of $Q$ is less than unity and
that the peak value of $Q$ in each ordered region decreases
monotonically with $I_0$. Note also that the increase of the
driving frequency $\Omega$ enhances the value of ${I_0}^{\left( 0
\right)}$ and moves the dynamically ordered regions to larger
values of $I_0$, which reflects that a higher frequency implies a
shorter period and accordingly, larger driving is necessary to
move the chirality lattice over the lattice barrier within the
(shorter) period. Except for the scale, however, the overall
behavior at zero temperature does not change qualitatively with
$\Omega$.

\section{Dynamic Transition}

\begin{figure}
\epsfig{width=0.4\textwidth,file=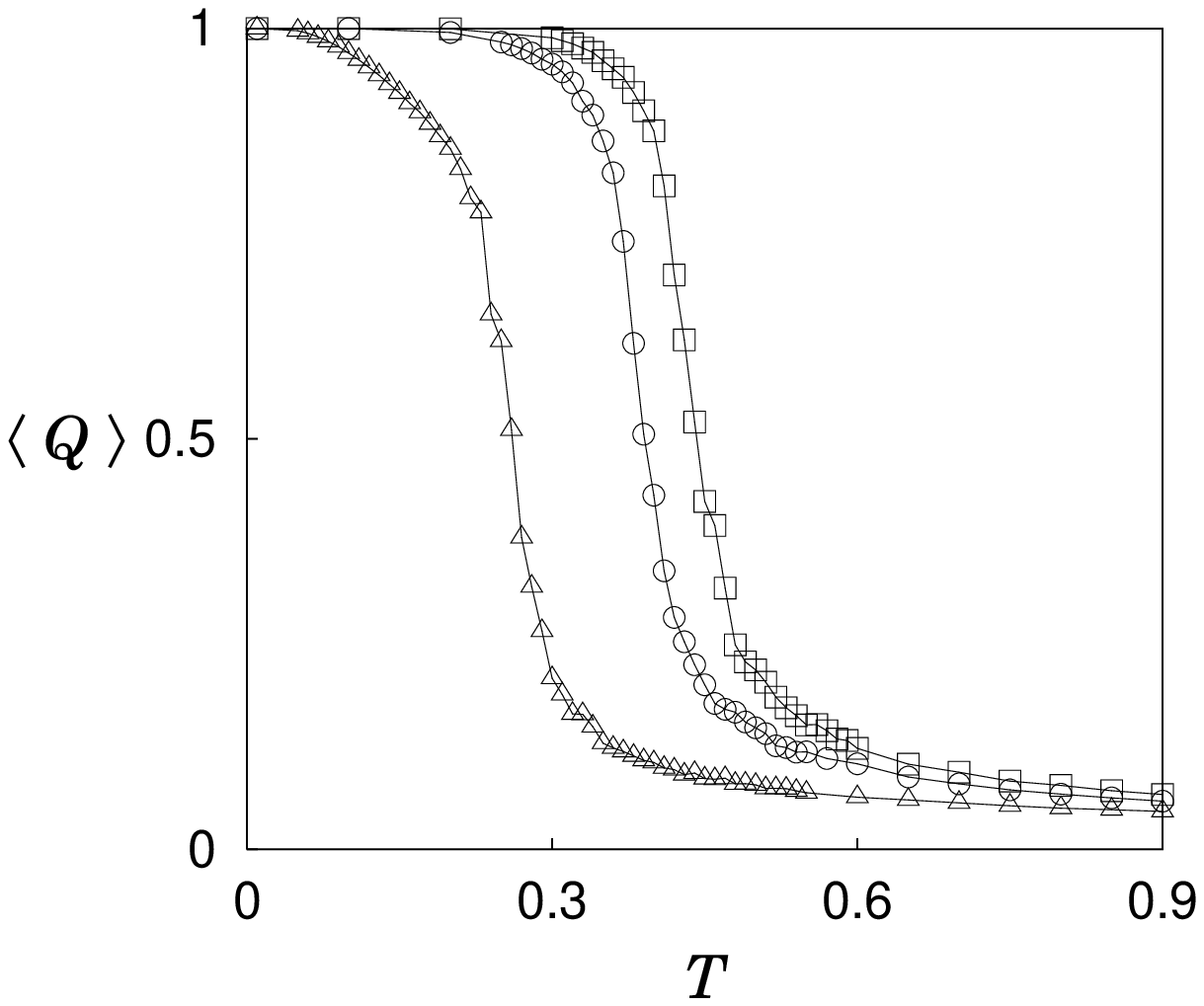} \\
\vspace{0.1cm}
(a) \\
\epsfig{width=0.4\textwidth,file=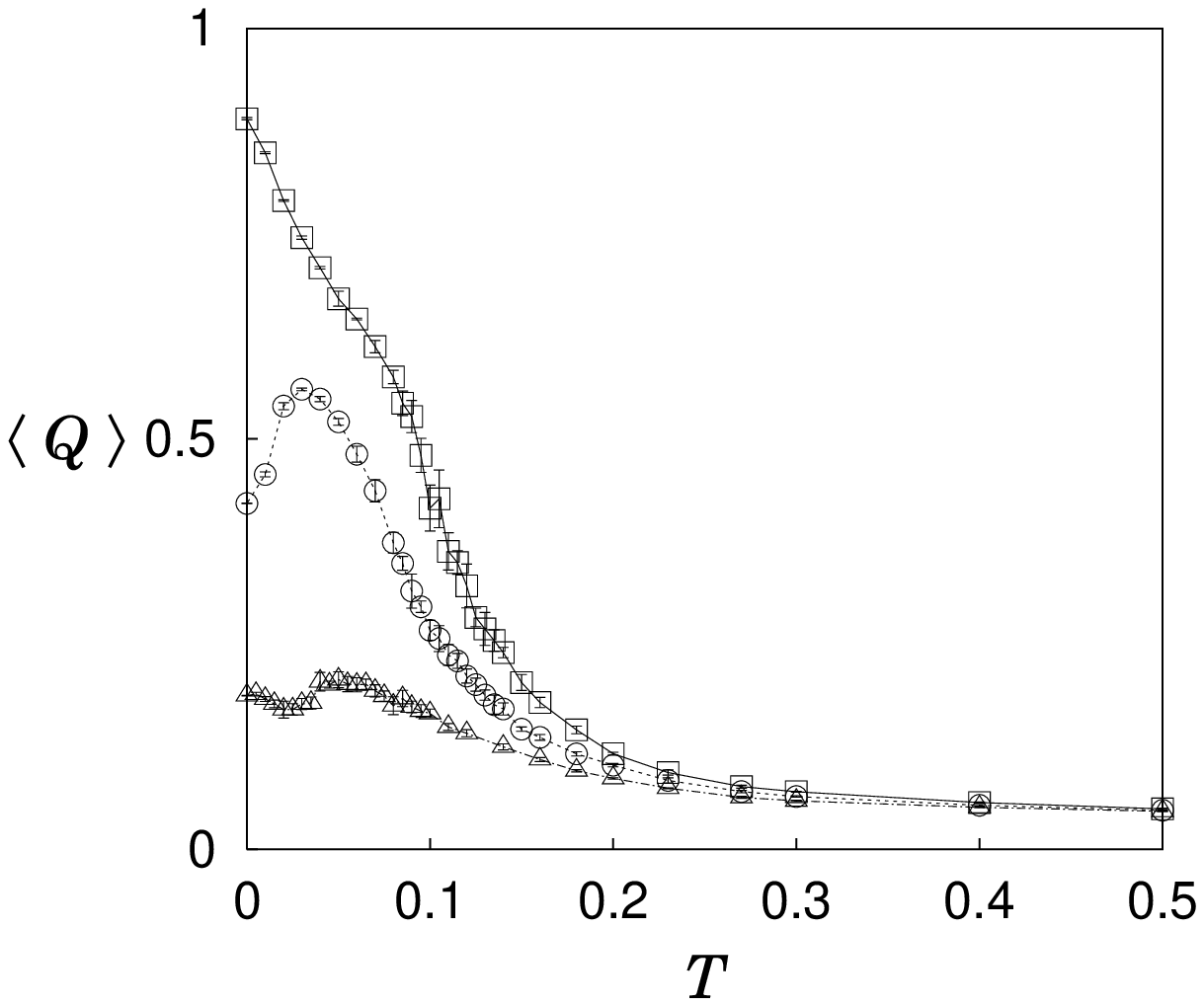} \\
\vspace{0.1cm} (b) \vspace{0.3cm} \caption[Dynamic order parameter
as a function of the temperature]{ Dynamic order parameter as a
function of the temperature in the system of size $L = 16$ for
various driving amplitudes (a) $I_0 = 0.3(\square), 0.5(\bigcirc),
0.8(\triangle)$; (b) $I_0 = 1.02(\square), 1.04(\bigcirc),
1.06(\triangle)$.  Error bars have been estimated from standard
deviations; in (a) they are not larger than the sizes of the
symbols.  Lines are merely guides to eyes.} \label{fig3}
\end{figure}

We next investigate the effects of the temperature.
At sufficiently high temperatures, thermal fluctuations are dominant so that
the influence of the driving current and the lattice pinning potential can be neglected.
Consequently, at high temperature, the staggered magnetization should fluctuate
randomly with time, resulting in the vanishing dynamic order parameter.
On the other hand, at low temperatures, where thermal fluctuations are small,
the system is expected to be driven mainly by the oscillating uniform currents,
exhibiting behavior similar to that at zero temperature.
Accordingly, we expect a phase transition between the dynamically ordered phase
and the disordered one as the temperature is varied.
Figure~\ref{fig3}, in which the ensemble average of the dynamic order parameter
$\langle Q \rangle$ is plotted as a function of the temperature $T$
for various driving amplitudes, explicitly shows the existence of such a transition
between the dynamically ordered phase and the disordered one.
The dynamic order parameter, starting from zero at high temperatures,
develops as the temperature is lowered.
It then grows rapidly and saturates eventually to the zero-temperature value
as the temperature approaches zero.
In particular, for $I_0$ above ${I_0}^{(0)}$, Fig.~\ref{fig3}(b) shows
that the zero-temperature value of $\langle Q \rangle$ is reduced rapidly
with the increase of $I_0$.
Here, it is interesting that the apparent peak of $\langle Q \rangle$ is observed
at a certain finite temperature rather than at zero temperature.

\begin{figure}
\centering \epsfig{width=0.4\textwidth,file=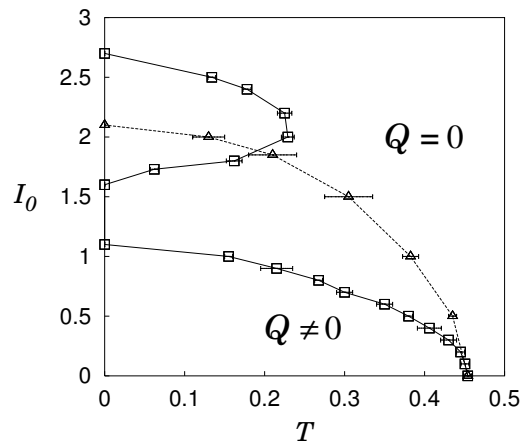}
\vspace{0.3cm} \caption[Dynamic phase diagram]{Dynamic phase
diagram on the $T{-}I_0$ plane for driving frequencies
$\Omega/2\pi = 0.08(\square)~ {\rm and ~} 0.16(\triangle)$. The
boundaries are determined by the crossing points of Binder's
cumulants for size $L = 8, 16$, and $24$.  Again error bars have
been estimated from standard deviations and lines are merely
guides to eyes.} \label{fig4}
\end{figure}

To estimate the transition temperature, we consider Binder's cumulants \cite{Binder}
\begin{equation}
U_L = 1 - \frac{\langle Q^4 \rangle}{3{\langle Q^2 \rangle}^2}.
\end{equation}
Since this quantity is size-independent at the transition temperature,
the transition temperature can be determined from the crossing point of $U_L$
for several sizes $L$.  In Fig.~\ref{fig4}, we plot the resulting
phase diagram on the $T{-}I_0$ plane, displaying dynamic phase boundaries.
The transition temperature initially decreases monotonically to zero as the driving
amplitude $I_0$ is increased.
However, further increase of $I_0$ drives the system into another ordered region, where
the transition temperature first grows with the driving current, then reduces to zero.
The zero-temperature results shown in Fig.~\ref{fig2} suggest that these additional
ordered regions emerge repeatedly with the amplitude $I_0$ raised
although the peak values of the transition temperature in these regions should
decrease with $I_0$.
As the driving frequency $\Omega$ is increased, on the other hand, the transition
temperature also increases and the dynamically ordered region expands
on the $T{-}I_0$ plane.
It is thus concluded that high driving frequency helps to maintain the
dynamic order, which is consistent with the zero-temperature result in
Fig.~\ref{fig2}.
It should also be noted here that cooling and heating curves for the dynamic order
parameter do not exhibit any appreciable hysteresis even in the strong-current regime.

\begin{figure}
\centering
\epsfig{width=0.4\textwidth,file=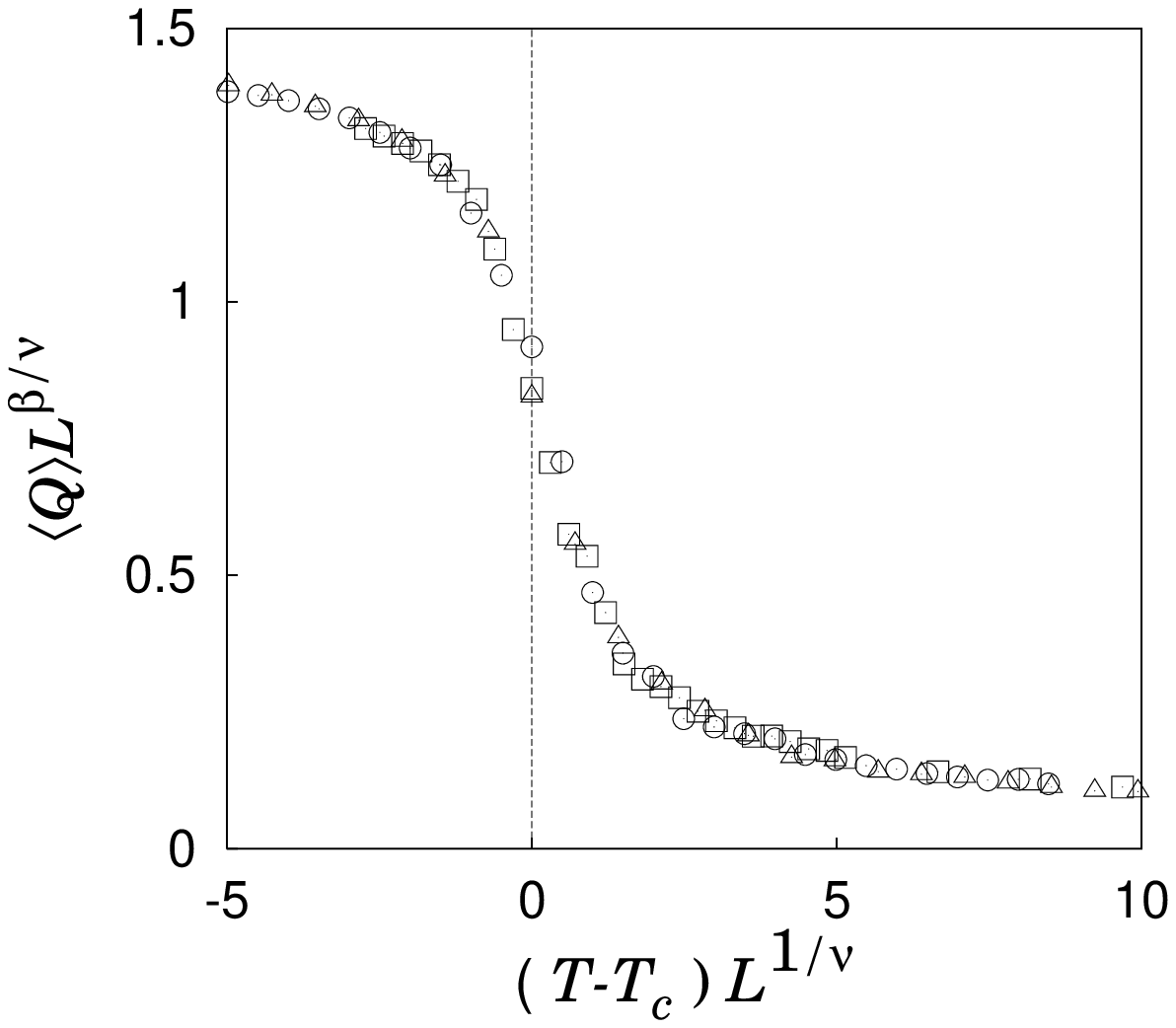} \\
\vspace{0.1cm}
 (a) \\
\centering
\epsfig{width=0.4\textwidth,file=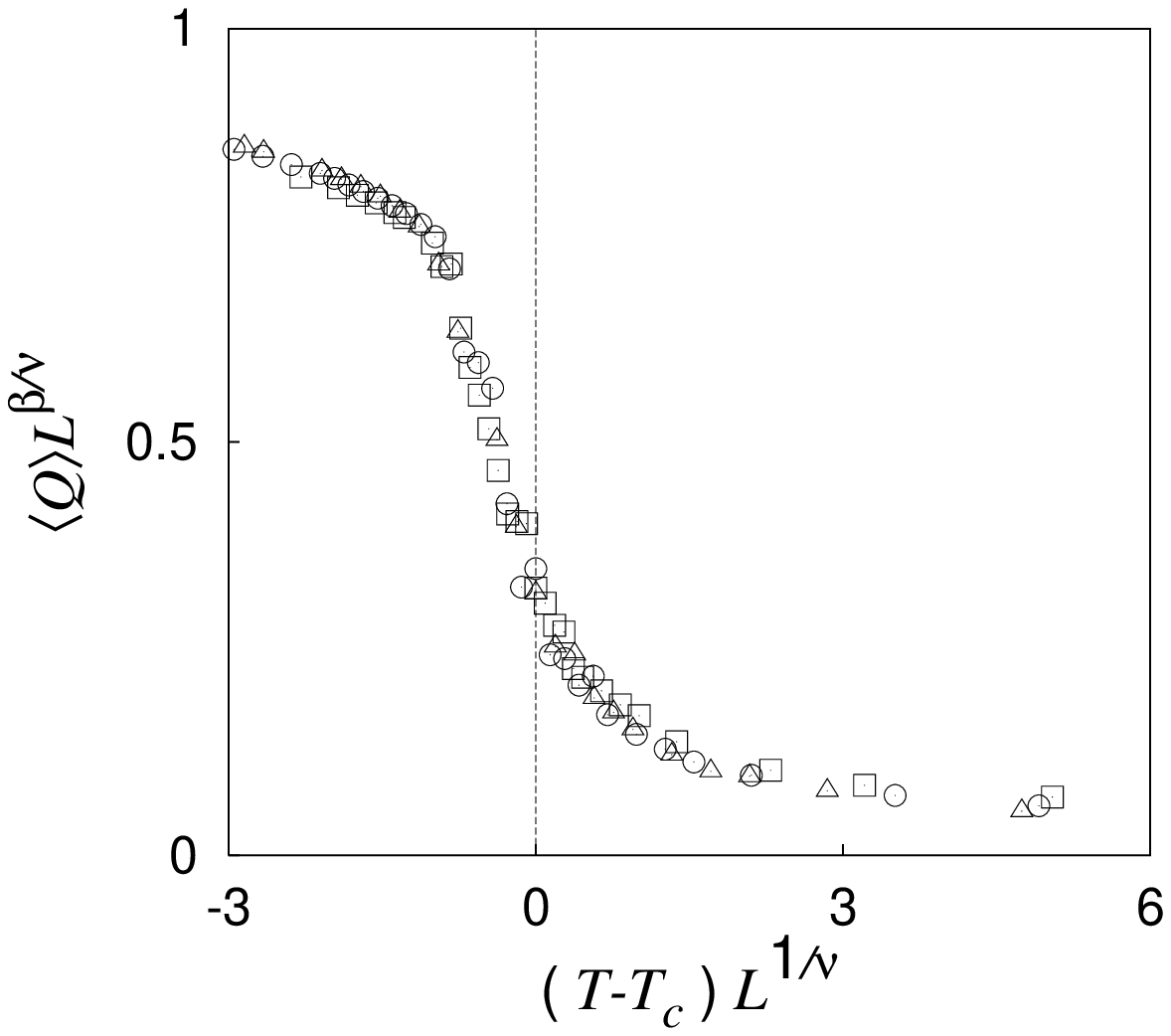} \\
\vspace{0.1cm}
 (b)
\vspace{0.3cm} \caption[Scaling plot of the dynamic parameter
versus the temperature] {Scaling plot of the dynamic order
parameter versus the temperature for size $L = 16(\square)$,
$24(\bigcirc)$, and $32(\triangle)$ with $\Omega/2\pi = 0.08$ and
(a) $I_0 = 0.3$; (b) $I_0 = 2.2$. The data fitting has been made
with the equilibrium values of the critical exponents (a) $\nu =
0.82$ and $\beta/\nu = 0.11$; (b) $\nu = 0.95$ and $\beta/\nu =
0.13$.} \label{fig5}
\end{figure}

To probe the nature of the transition, we consider the scaling relation
for the dynamic order parameter
\begin{equation}
\langle Q \rangle = L^{-\beta/\nu} f\left((T - T_c)L^{1/\nu}\right)
\end{equation}
and plot $\langle Q \rangle L^{\beta/\nu}$ versus $(T -
T_c)L^{1/\nu}$ for various sizes. Figure~\ref{fig5} exhibits the
resulting scaling plots for the driving amplitude (a) $I_0 = 0.3$
and (b) $I_0 = 2.2$, with the critical exponents: (a) $\nu = 0.82$
and $\beta/\nu = 0.11$; (b) $\nu = 0.95$ and $\beta/\nu = 0.13$.
Since the critical exponents in (a) agree well with those for the
equilibrium ${\rm Z}_2$ transition in the FF$XY$
model,\cite{luo,slee} the nice collapse of the data suggests
strongly that the dynamic transition in the first ordered region
[i.e., $I_0 < I_0^{(0)}$ in Fig.~\ref{fig2}(a)] belongs to the
same universality class as the equilibrium ${\rm Z}_2$ transition
of the FF$XY$ model.  Similar conclusion was also reached in the
FFJJA under weak staggered oscillating fields.\cite{gun} On the
other hand, the values in (b), corresponding to the second ordered
region, appear somewhat larger than the critical exponents for the
equilibrium ${\rm Z}_2$ transition. However, due to large
fluctuations, the numerical accuracy in (b) is not so good,
yielding reasonable collapsing behavior down to $\nu = 0.87$ and
$\beta/\nu = 0.12$. Further, note that there does exist rather
large discrepancy among the equilibrium values reported in
literature,\cite{luo,slee,exponent} which may be summed up as
follows: $\beta = 0.11 \pm 0.03$ and $\nu = 0.9 \pm 0.1$. In view
of this uncertainty, it is also possible to consider the values in
(b) to be consistent with those for the same (equilibrium ${\rm
Z}_2$) transition.
%
Accordingly, we presume that for all driving amplitudes the dynamic transition
in the system belongs to the same universality class as the equilibrium Z$_2$
transition present in the FF$XY$ model.

\section{Resonance Behavior}

Finally, we explore the possibility of the stochastic resonance phenomena
in the FFJJA under uniform oscillating currents.
In numerical simulations, we compute the power spectrum $P(\omega)$
of the staggered magnetization $m(t)$ and its SNR, which is defined to be
\begin{equation}
{\rm SNR} = 10 \log_{\rm 10} \left[ \frac{S}{N} \right].
\end{equation}
The signal $S$ is given by the peak intensity of the power spectrum
at the driving frequency $\Omega$ and $N$ represents the the background noise level,
which is estimated by the average power spectrum around the signal peak.
We can also define similar quantities for the higher harmonics of
the driving frequency $\Omega$, the behaviors of which are found to be
quite peculiar as described below.

\begin{figure}
\centering
\epsfig{width=0.4\textwidth,file=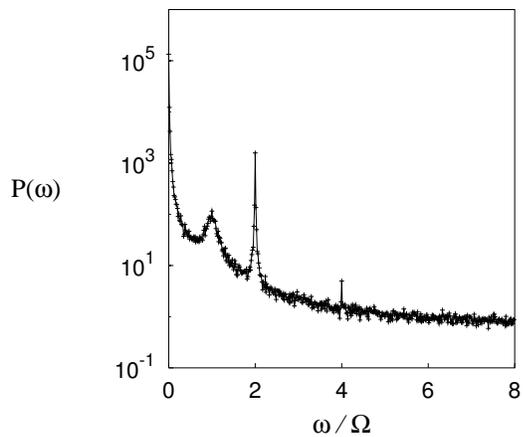} \\
\vspace{0.1cm}
 (a) \\
\epsfig{width=0.4\textwidth,file=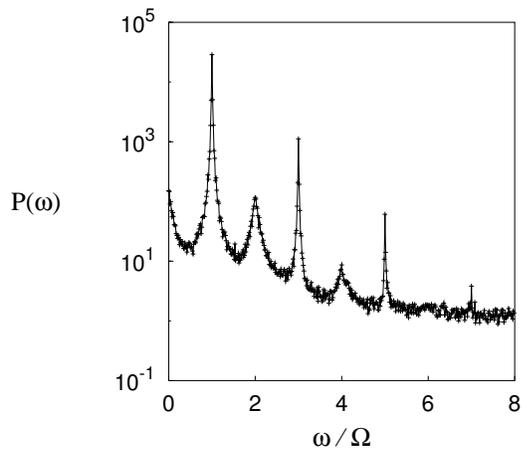} \\
\vspace{0.1cm}
 (b) \\
\epsfig{width=0.4\textwidth,file=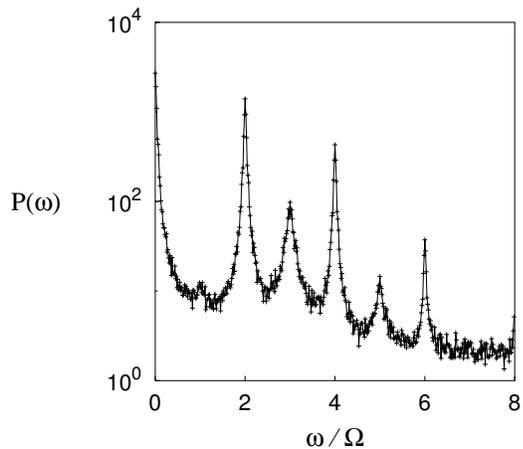} \\
\vspace{0.1cm}
(c)
\vspace{0.3cm}
\caption[Power spectrum of the staggered magnetization]
{Power spectrum of the staggered magnetization for size $L = 16$ and
driving frequency $\Omega/2\pi = 0.08$.  The driving amplitude and the temperature
are (a) $I_0 = 0.8$ and $T = 0.28$; (b) $I_0 = 1.2$ and $T = 0.2$;
(c) $I_0 = 2.0$ and $T = 0.3$.}
\label{power}
\end{figure}

In Fig.~\ref{power}, we display typical behavior of the power spectrum $P(\omega)$
at low temperatures for several driving amplitudes.
Note the substantial difference according to the driving current amplitude $I_0$,
which may be understood in view of zero-temperature states.
For $I_0 = 0.8$, where there is no oscillation of the staggered magnetization $m(t)$
at zero temperature, observed at low temperatures are broad peaks around
odd harmonics as well as sharp peaks at even harmonics, as shown in Fig.~\ref{power}(a).
On the other hand, for higher values of $I_0$, corresponding to $Q = 0$
at zero temperature, addition of thermal noise gradually suppresses the sharp peaks at
odd harmonics while inducing broad peaks at even harmonics [see Fig.~\ref{power}(b)].
When the system enters another ordered region of $0 < Q < 1$ by further increase
of $I_0$, the staggered magnetization exhibits only oscillations of even harmonics
at zero temperature, which remain as sharp peaks at finite temperatures.
Here, noise again helps to develop broad peaks around odd harmonics and interestingly,
the third harmonics are dominant over the first harmonics [see Fig.~\ref{power}(c)].

\begin{figure}
\centering
\epsfig{width=0.4\textwidth,file=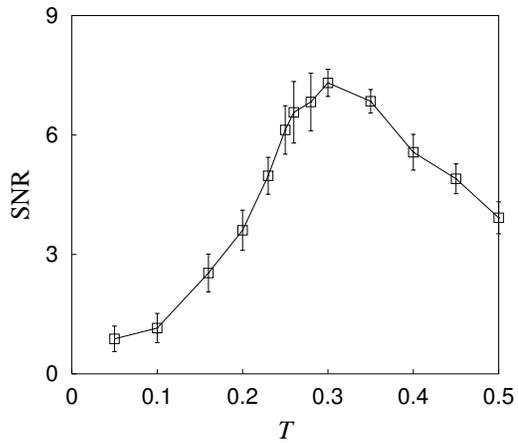} \\
\vspace{0.1cm}
(a) \\
\epsfig{width=0.4\textwidth,file=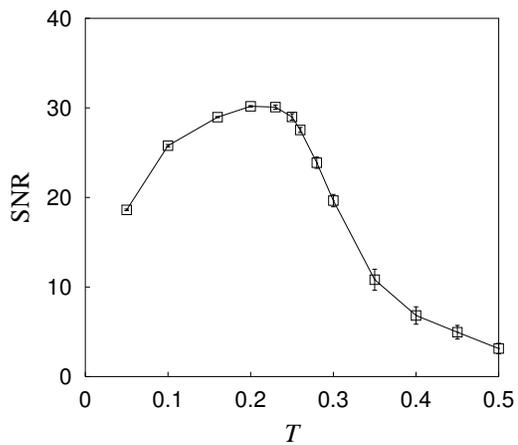} \\
\vspace{0.1cm}
(b)
\vspace{0.3cm}
\caption[Signal-to-noise ratio versus the temperature for the driving amplitude
$I_0 = 0.8$]
{Signal-to-noise ratio versus the temperature in the system of size $L = 16$
with $\Omega/2\pi = 0.08$ and $I_0 = 0.8$.  The signal-to-noise ratio is calculated
(a) at the first harmonics and (b) at the second harmonics.
Error bars have been estimated from standard deviations;
lines are guides to eyes.}
\label{fig7}
\end{figure}

\begin{figure}
\centering
\epsfig{width=0.4\textwidth,file=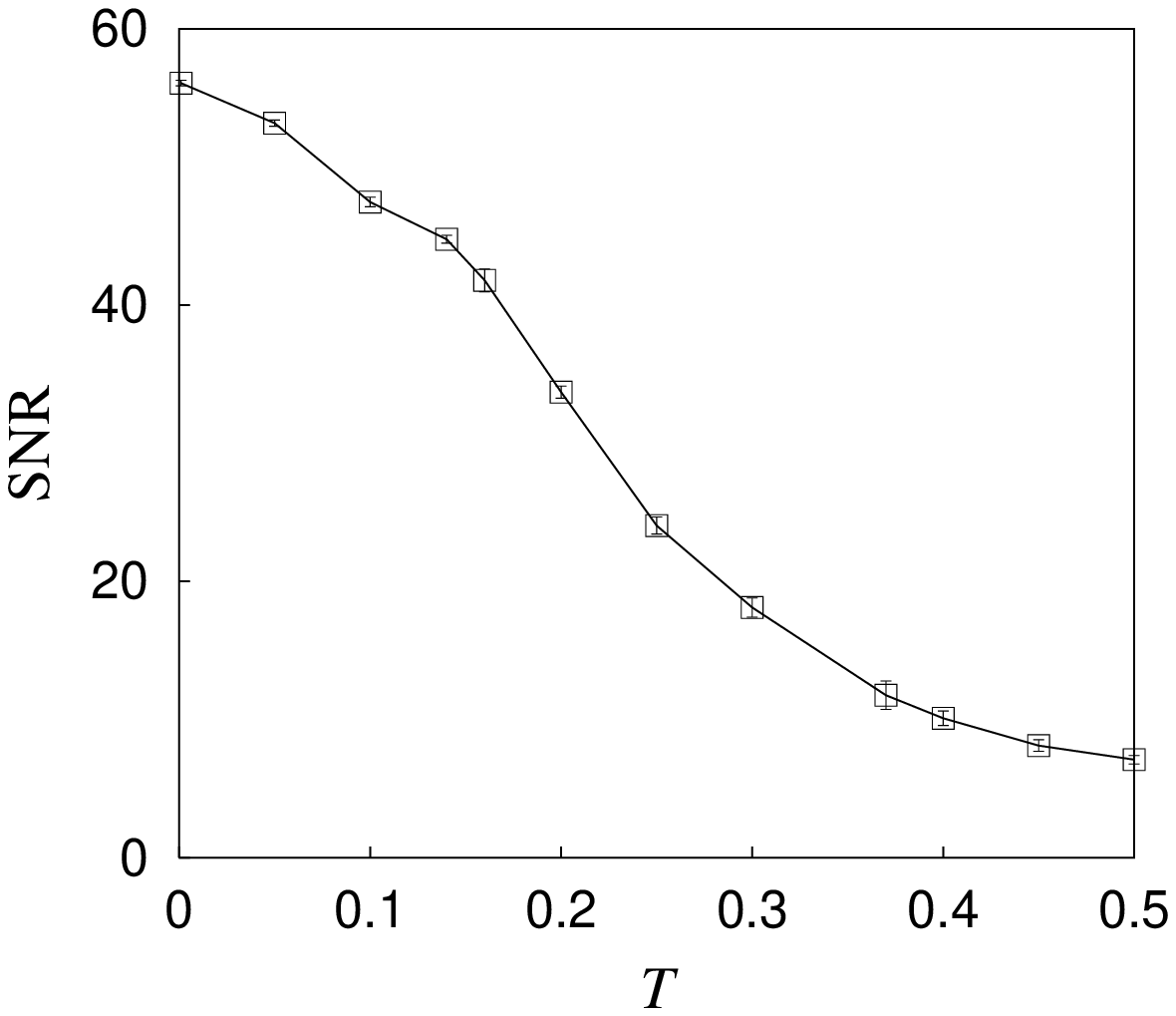} \\
\vspace{0.1cm}
(a) \\
\epsfig{width=0.4\textwidth,file=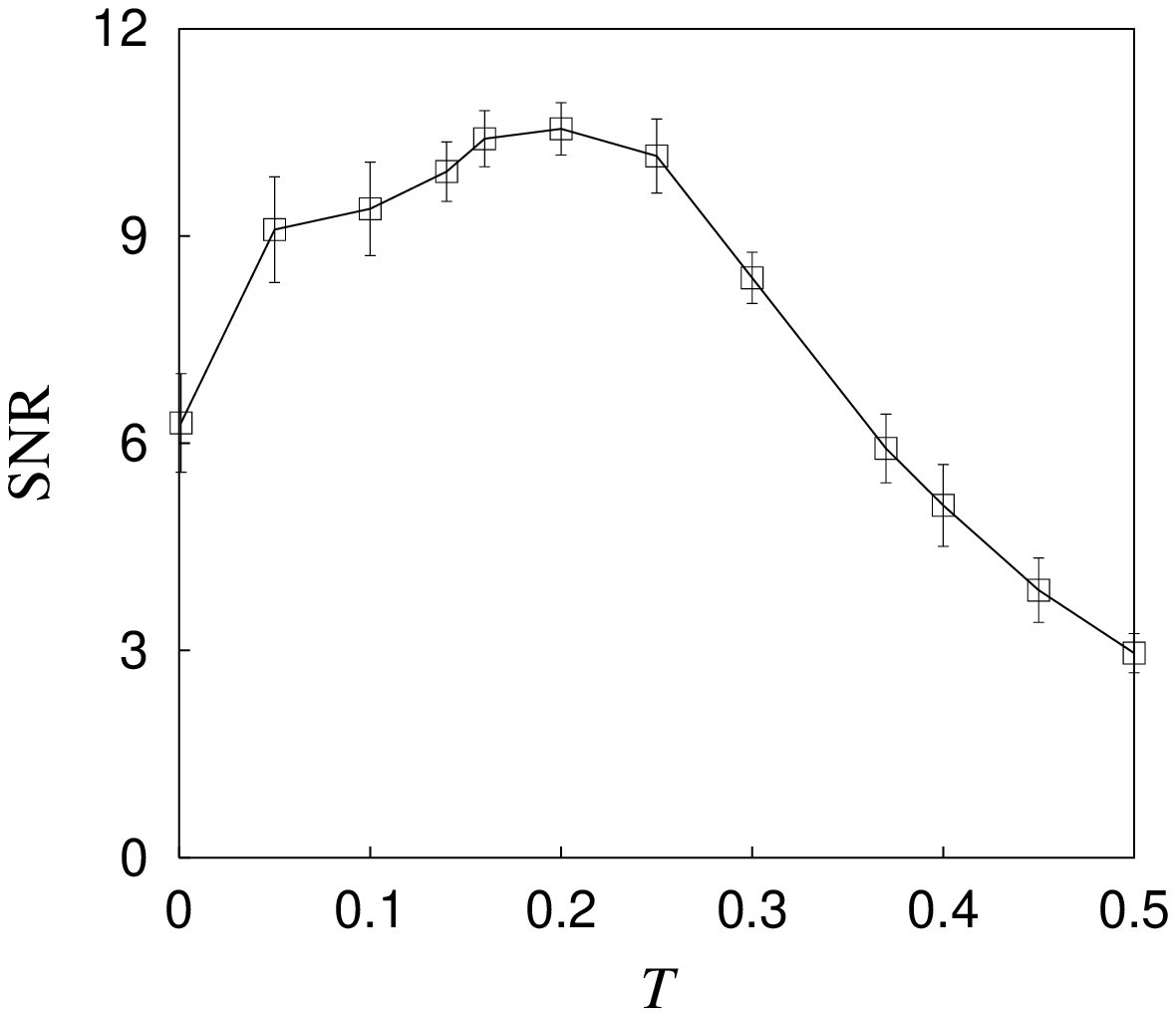} \\
\vspace{0.1cm}
(b)
\vspace{0.3cm}
\caption[Signal-to-noise ratio versus the temperature for the driving amplitude
$I_0 = 1.2$]
{Signal-to-noise ratio versus the temperature in the system of size $L = 16$
with $\Omega/2\pi = 0.08$ and $I_0 = 1.2$ The signal-to-noise ratio is calculated
(a) at the first harmonics and (b) at the second harmonics.
The leftmost data correspond to the temperature $T = 0.001$.}
\label{fig8}
\end{figure}

\begin{figure}
\centering
\epsfig{width=0.4\textwidth,file=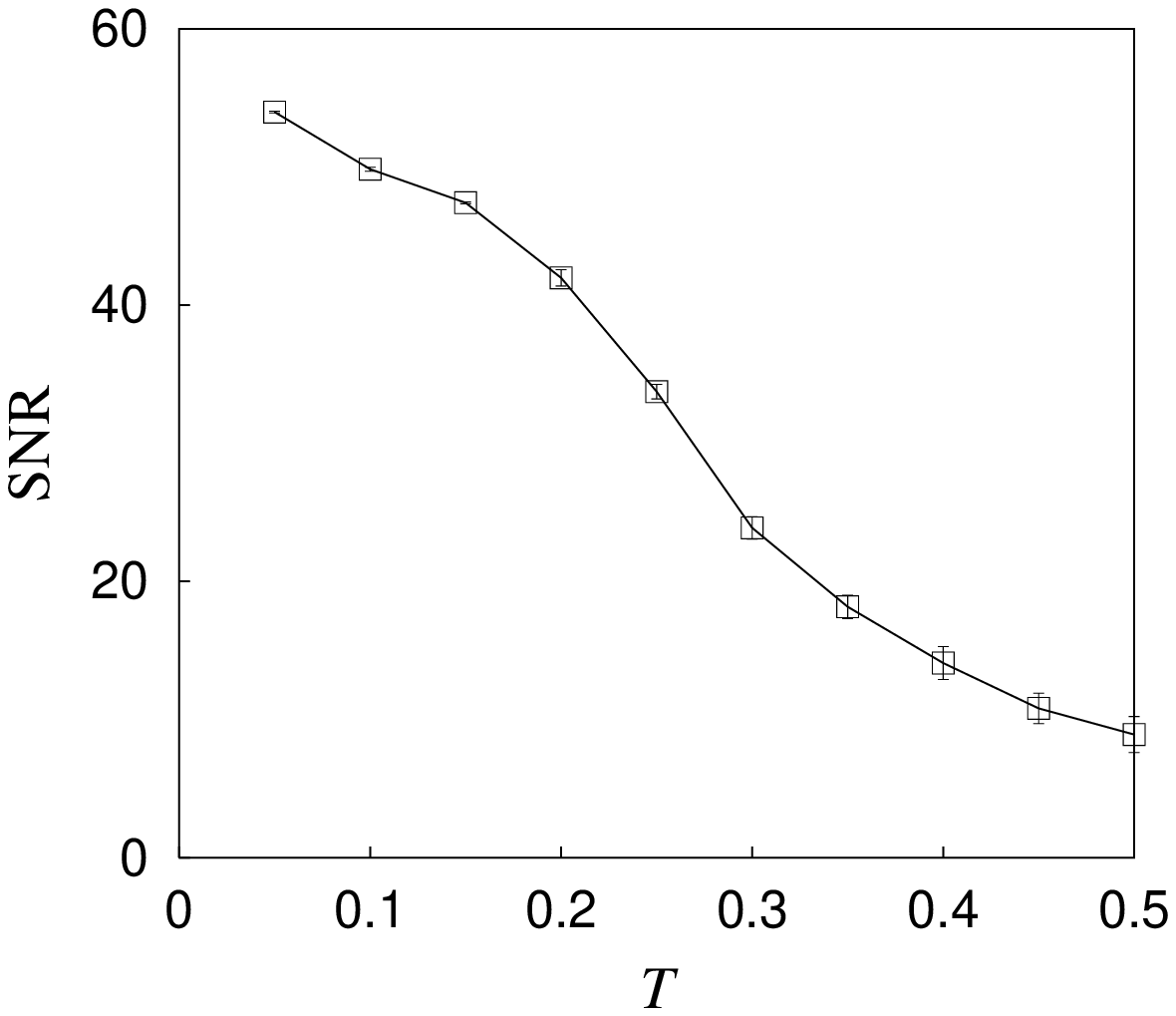} \\
\vspace{0.1cm}
(a) \\
\epsfig{width=0.4\textwidth,file=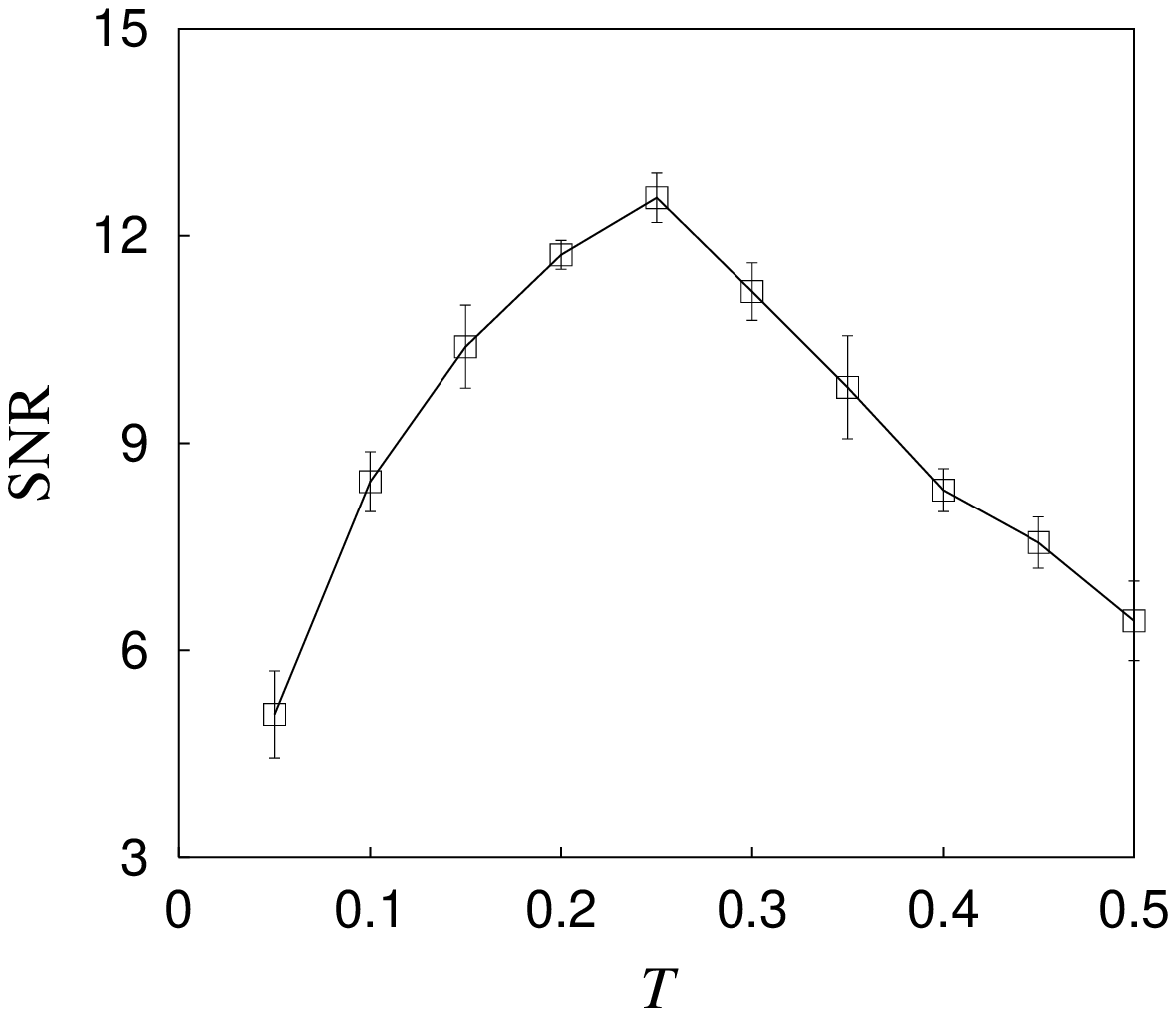} \\
\vspace{0.1cm}
(b)
\vspace{0.3cm}
\caption[Signal-to-noise ratio versus the temperature for the driving amplitude
$I_0 = 2.0$]
{Signal-to-noise ratio versus the temperature in the system of size $L = 16$
with $\Omega/2\pi = 0.08$ and $I_0 = 2.0$  The signal-to-noise ratio is calculated
(a) at the second harmonics and (b) at the third harmonics.}
\label{fig9}
\end{figure}

Such interesting behavior depending on the driving current
amplitude is expected to bring about rich physics in the SR
phenomena.  The SNR computed at the first and second harmonics for
$I_0 = 0.8$ is plotted in Fig.~\ref{fig7} (a) and (b),
respectively. It is demonstrated that both the broad peak at the
first harmonics and the sharp one at the second harmonics, which
shows up in the presence of the noise, exhibit the SR behavior.
Remarkably, the resonance temperature for the first harmonics is
found to be different from that for the second harmonics: The
former is higher and the dynamic transition temperature is located
between the two resonance temperatures.  For $I_0 = 1.2$, the SR
behavior of the second harmonics is again observed as shown in
Fig.~\ref{fig8}(b). However, the SNR of the first harmonics
decreases monotonically with the temperature [see
Fig.~\ref{fig8}(a)], indicating that the oscillations of the odd
harmonics present in the zero-temperature state are simply
suppressed by the noise. Similar argument explains the SR
behaviors for $I_0 = 2.0$, where only the even harmonics exist in
the oscillations of the staggered magnetization at zero
temperature. The thermal noise tends to reduce the oscillations of
the second harmonics, which is manifested by the monotonic
decrease of the SNR in Fig.~\ref{fig9}(a). In contrast, the SNR of
the third harmonics is initially enhanced by the weak thermal
noise as shown in Fig.~\ref{fig9}(b). These interesting behaviors
may be understood as follows: The thermal noise, which is not
strong enough to destroy the chirality lattice, is expected to
help the chirality lattice move over one more lattice barrier if
the threshold is not so large. Accordingly, for $I_0=1.2$
corresponding to $Q=0$ at zero temperature, the even harmonics
which are absent in the ground state are induced by the thermal
noise while the odd harmonics are generally suppressed. Via the
same argument, one can expect that for $I_0=2.0$ in the second
ordered region, the odd harmonics exhibit the SR behavior with the
even harmonics decreasing monotonically. At higher values of
$I_0$, similar SR behavior is expected for odd and even harmonics
according to the zero-temperature oscillations of the staggered
magnetization.

\section{Conclusion}

We have examined the dynamic properties of the 2D FFJJA driven uniformly by
alternating currents, with attention paid to the dynamic transition
and the SR phenomena.
At low temperatures the chirality in the FFJJA displays antiferromagnetic ordering.
To describe the dynamic transition associated with the antiferromagnetic
ordering of the chirality, we have introduced the dynamic order parameter,
which is the staggered magnetization averaged over one period of
the oscillating current, and investigated its behavior with respect to the amplitude
and the frequency of the driving current as well as to the temperature.
At zero temperature, as the driving amplitude is raised, the dynamically ordered phase
and the disordered one have been observed to appear alternatively.
The saturation value of the dynamic order parameter is unity in the first ordered
region while it is less than unity in other ordered regions.
It has also been observed that the increase of the driving frequency tends to shift
the dynamically ordered regions toward larger values of the driving amplitude.
Except for the scale, the overall behavior at zero temperature does not change
qualitatively with the driving frequency.
As the temperature is increased, the ordered phase undergoes a dynamic phase
transition to the disordered phase.
We have obtained the phase diagram of the dynamic transition on the plane of the
temperature and the driving amplitude.
Through the use of the finite-size scaling analysis,
the dynamic transition has been shown to belong to the same universality class
as the equilibrium Z$_2$ transition in the FF$XY$ model.
We have also examined the SR phenomena in the system.
To investigate the phenomena, we have calculated the power spectrum of
the staggered magnetization, which shows characteristic behavior
according to the state at zero temperature.
In the regions corresponding to finite values of the dynamic order parameter
at zero temperature, broad peaks have been observed around odd harmonics
as well as sharp peaks at even harmonics.
On the other hand, in the regions with vanishing dynamic order parameters
at zero temperature, the opposite situation has been observed.
As manifested by the SNR at several harmonics,
the odd and the even harmonics have been revealed to exhibit different SR behaviors,
which may be understood in view of the variation of the zero-temperature states
with the driving amplitude.

\begin{acknowledgments}
MYC thanks the Korea Institute for Advanced Study for hospitality during his visit,
where part of this work was accomplished.
This work was supported in part by
the Ministry of Education through the BK21 Program
and in part by the Korea Science and Engineering Foundation through the Center
for Strongly Correlated Materials Research (GSJ).
\end{acknowledgments}

\end{document}